\documentstyle[emulateapj,onecolfloat]{article}

\input epsf

\newcommand{\beq}{\begin{equation}}
\newcommand{\eeq}{\end{equation}}

\newcommand{\gsim}{\gtrsim}

\newcommand{\Omegat}{\Omega_{\rm t}}
\newcommand{\wm}{\omega_m}
\newcommand{\wb}{\omega_b}

\begin{document}
\twocolumn[
\title{CMB Observables and Their Cosmological Implications}

\author{Wayne Hu,$^1$\altaffilmark{4} Masataka Fukugita,$^{1,2}$ Matias Zaldarriaga,$^1$ and Max Tegmark$^3$}
\affil{{}$^1$ Institute for Advanced Study, Princeton, NJ 08540, USA,
\\ 
       {}$^2$ Institute for Cosmic Ray Research, University of Tokyo,
Tanashi, Tokyo 188, Japan \\
       {}$^3$ Department of Physics, University of Pennsylvania, 
	PA, 19104      
}

\begin{abstract}
We show that recent measurements of the power spectrum of 
cosmic microwave background anisotropies by BOOMERanG and MAXIMA 
can be mainly characterized by 
four observables, the position of the first acoustic peak $\ell_1
= 206 \pm 6$,
the height of the first peak relative to COBE normalization 
$H_1= 7.6 \pm 1.4$, 
the height of the second peak relative to the first 
$H_2 = 0.38 \pm 0.04$,
and the height of the third peak relative to the 
first $H_3 = 0.43 \pm 0.07$.
This phenomenological representation of the measurements
complements more detailed likelihood analyses in 
multidimensional parameter space, clarifying
the dependence on prior assumptions and 
the specific aspects of the data leading to the constraints.
We illustrate their use in the flat $\Lambda$CDM family of models
where we find $\Omega_m h^{3.8} > 0.079$
(or nearly equivalently, the age of the universe $t_0 < 13-14$Gyr) 
from $\ell_1$ and a baryon 
density $\Omega_b h^2 > 0.019$, a 
matter density $\Omega_m h^2 < 0.42$ 
and tilt $n>0.85$ from the peak heights (95\% CL). 
With the aid of several external constraints, notably nucleosynthesis,
the age of the universe and the cluster abundance and baryon fraction, we construct the allowed region in the ($\Omega_m$,$h$) plane; 
it points to high $h$ ($0.6<  h < 0.9$)
and moderate $\Omega_m$ ($0.25 < \Omega_m  < 0.6$).
\end{abstract}
]

\altaffiltext{4}{Alfred P. Sloan Fellow}
\section{Introduction}

With the data from the BOOMERanG 
(\cite{deBetal00} 2000) and MAXIMA (\cite{Hanetal00} 2000) 
experiments, the promise of measuring cosmological parameters
from the power spectrum of anisotropies in the cosmic 
microwave background (CMB) has come substantially closer
to being fulfilled.
Together they determine the location of the first peak precisely and
constrain the amplitude
of the power at the expected position of the second peak. 
The MAXIMA experiment also 
limits the power around the expected rise
to the third peak.  
%It has already been shown that these observations 
%strongly constrain cosmological parameters (\cite{Lanetal00} 2000;
%\cite{Baletal00} 2000; \cite{TegZal00b} 2000b). 
%Whether these constraints agree with those 
%from other cosmological observations serves as a fundamental
%test of the underlying 
%adiabatic cold dark matter (CDM) model of structure formation.

These observations strongly constrain cosmological parameters 
as has been shown through 
likelihood analyses
in multidimensional parameter space with a variety of prior assumptions
(\cite{Lanetal00} 2000; \cite{Baletal00} 2000;
\cite{TegZal00b} 2000b; \cite{Brietal00} 2000).
While these analyses are complete in and of themselves, the 
high dimension of the parameter space makes it difficult to 
understand what characteristics 
of the observations or prior assumptions
are driving the constraints. 
For instance, it has been claimed that the BOOMERanG data favor
closed universes (\cite{Whietal00} 2000; \cite{Lanetal00} 2000)
and high baryon density (\cite{Hu00} 2000; \cite{TegZal00b} 2000b),
but the role of priors, notably from the Hubble constant
and big-bang nucleosynthesis is less clear. 
Indeed, whether CMB constraints agree with those 
from other cosmological observations serves as a fundamental
test of the underlying 
adiabatic cold dark matter (CDM) model of structure formation.

In this paper we show that most of the information in the power
spectrum from these two data sets can be compressed into four 
observables.  The correlation among cosmological parameters 
can be understood by 
studying their effects on the four observables.  
They can also be used to search for solutions outside the standard
model space (e.g. 
\cite{PeeSeaHu00} 2000; 
\cite{Bouetal00} 2000).

As an instructive application of this approach, we consider 
the space of flat adiabatic CDM models.
Approximate flatness is clearly favored 
by both BOOMERanG and MAXIMA (\cite{deBetal00} 2000;
\cite{Hanetal00} 2000) as well as previous data, notably from
the TOCO experiment (\cite{Miletal99} 1999), as shown by
previous analyses (\cite{Lin98} 1998; \cite{Efsetal99} 1999; \cite{TegZal00a} 2000a).

\begin{figure*}[tbh]
\centerline{\epsfxsize=7.25truein\epsffile{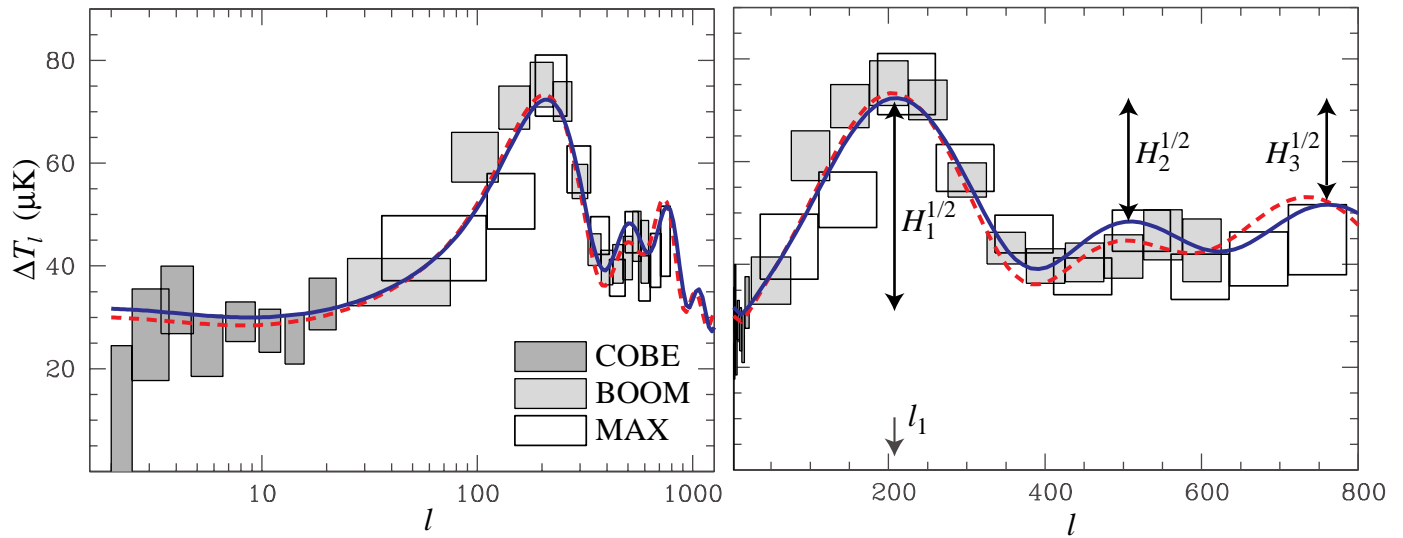}}
\caption{\footnotesize Power spectrum data and models: (left panel) full range on
a log scale; (right panel) first 3 peaks on a linear scale.  
The BOOMERanG (BOOM) and MAXIMA (MAX) points have been shifted by their 1$\sigma$ calibration
errors, $10\%$ up and $4\%$ down respectively. 
Dashed lines represent 
a model that is a good fit to the CMB data alone:  
$\Omega_m=0.3$, $\Omega_\Lambda=0.7$, $h=0.9$, $\Omega_b h^2=0.03$, $n=1$ which gives 
$\ell_1= 205$, $H_1=6.6$, $H_2=0.37$,  $H_3=0.52$.  
Solid lines represent a model that is allowed by our joint constraints:
$\Omega_m=0.35$, $\Omega_\Lambda=0.65$, $h=0.75$, $\Omega_b h^2=0.023$, $n=0.95$
which gives $\ell_1= 209$, $H_1=5.8$, $H_2=0.45$, $H_3=0.5$.  Note that
the labelling of the $H$'s in the figure is schematic; these values are the
power {\it ratios} as defined in the text.
}
\label{fig:datamodels}
\end{figure*}

Our main objective in this application is to clarify the constraints
derived from the CMB observations using the likelihood analyses and 
understand how they might change as the data evolves.
Then, with the aid of a few external constraints 
we map out the allowed region in 
the plane of the matter density ($\Omega_m$) 
versus the Hubble constant ($H_0$: we use $h$ to denote the 
Hubble constant $H_0=100h$ km s$^{-1}$Mpc$^{-1}$). 
The external constraints which we employ include
(i) the rich cluster abundance at $z\approx 0$, (ii) the cluster baryon
fraction, (iii) the baryon abundance from big bang
nucleosynthesis (BBN), and (iv) the minimum age of the universe. 
We also discuss their consistency with 
other constraints, such as direct 
determinations of $H_0$, $\Omega_m$, and the luminosity 
distance to high redshift supernovae.
All errors we quote in this paper are at 67\% confidence, 
but we consider all constraints at a
95\% confidence level. 

In \S 2 we start with a statistical analysis of the CMB data.  
We introduce the four observables and discuss
their cosmological implications.
In \S 3, we place constraints on the $(\Omega_m,h)$ plane
and discuss consistency checks.  
In \S 4, we identify opportunities for future consistency 
checks and arenas for future confrontations with data.
We conclude in \S 5.
The appendix presents convenient formulae
that quantify the cosmological parameter dependence of our four
characteristic observables in adiabatic CDM models. 
 
\section{CMB Observables}

\subsection{Statistical Tests}

With the present precision of the BOOMERanG and MAXIMA observations 
(see Fig.~\ref{fig:datamodels}),
it is appropriate to characterize the power spectrum with
four numbers:
the position of the first peak $\ell_1$,
the height of the first peak relative to the power at $\ell=10$
\begin{equation}
H_1 \equiv \left({\Delta T_{\ell_1} \over  \Delta T_{10}}\right)^2\,,
\end{equation}
the height of the second peak relative to the first
\begin{equation}
H_2 \equiv \left({\Delta T_{\ell_2} \over  \Delta T_{\ell_1}}\right)^2\,,
\end{equation}
and the height of the third peak relative to the first
\begin{equation}
H_3 \equiv \left({\Delta T_{\ell_3} \over  \Delta T_{\ell_1}}\right)^2\,,
\end{equation}
where $(\Delta T_{\ell})^2 \equiv \ell(\ell+1) C_\ell/2\pi$
with $C_\ell$ the power spectrum of the multipole moments of the
temperature field.
Note that the locations of the second and third peaks are set by their
harmonic relation to the first peak [see Appendix, eq. (\ref{eqn:harmonicseries})] 
and so 
$H_2$ and $H_3$ are well-defined even in the absence of clear detections
of the secondary peaks.

One could imagine two different approaches towards measuring these
four numbers. We could extract them using some form of parametrized 
fit such as a parabolic fit to the data (\cite{KnoPag00} 2000;
\cite{deBetal00} 2000). Alternately, we could
use template CDM models as calculated by CMBFAST 
(\cite{SelZal96} 1996),
and label them by the values of the four observables. 
We can measure $\chi^2$ for these CDM models and interpret them as
constraints in the four
observables. Both of these methods give similar results. We chose
the second one because it is more stable to changes in the 
$\ell$ ranges taken to correspond to each peak and incorporates the 
correct shape of the power spectra for CDM-like models. 

\begin{figure}[htb]
\centerline{\epsfxsize=3.4truein\epsffile{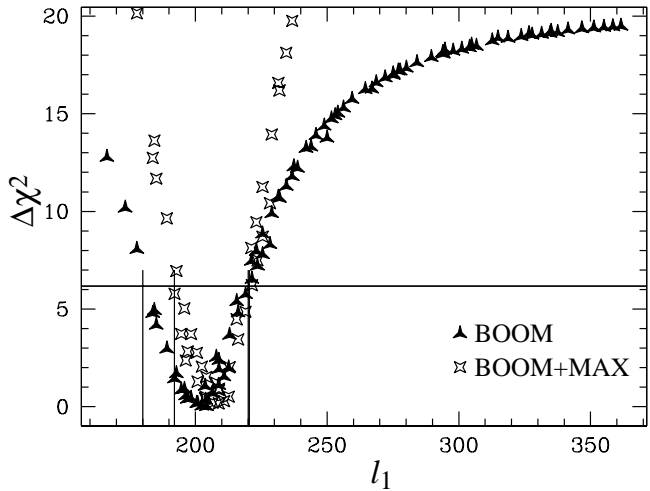}}
\caption{\footnotesize Constraints on the first peak position: 
$\Delta \chi^2(\ell_1)$ for the data from $75 < \ell < 375$.
We define the $1 \sigma$ errors to be $1/2.5$ of the errors
at $2.5 \sigma$ ($\Delta \chi^2=6.2$, solid lines).}
\label{fig:l1}
\end{figure}

To determine the position of the first peak, we take the data that
fall between $75 < \ell < 375$ and carry out a $\chi^2$ fitting
using a flat model template ($\Omega_{\rm m}+\Omega_\Lambda=1$ here
and below unless otherwise stated) with varying $h$ and $\Omega_m$ 
at the fixed baryon density $\Omega_bh^2=0.02$ and  
tilt parameter $n=1$.
We include calibration errors, 10\% for BOOMERanG
and 4\% for MAXIMA. 
Figure~\ref{fig:l1} shows $\Delta \chi^2$ as a function of $\ell_1$
for the BOOMERanG data alone and for the combination of BOOMERanG
and MAXIMA.  The figure implies
\begin{eqnarray}
\ell_1 &=& 200 \pm 8 \quad ({\rm BOOM}) \,,\nonumber\\
\ell_1 &=& 206 \pm 6 \quad ({\rm BOOM+MAX}) \,.
\label{eq:1}
\end{eqnarray}
Other choices of $\Omega_b h^2$ and $n$ for the template parameters
modify slightly the value of $\chi^2$ but not $\Delta\chi^2$ or
the allowed region for $\ell_1$.  It is noteworthy that adding 
in the MAXIMA data
steepens $\Delta\chi^2$ on {\it both} sides of the minimum despite
the preference for $\ell_1 \sim 220$ in the MAXIMA data alone
(\cite{Hanetal00} 2000).   The fact that both data sets consistently indicate 
a sharp fall in power at $\ell > 220$ increases the confidence 
level at which a high $\ell_1$ can be rejected.

The $H_1$ statistic depends both on the 
acoustic physics that determines the first peak and
other processes relevant at
$\ell \sim 10$. The shape of the
template is therefore more susceptible to model parameters.
We choose to vary $n$ which changes $H_1$ and 
also allow variations in $\Omega_\Lambda$ 
so that the position of the first peak can be properly adjusted. 
The other parameters were chosen to be 
$\Omega_b h^2=0.03$ 
and $\Omega_m h^2=0.2$.
Using the BOOMERanG and MAXIMA data for $75 <\ell<
375$ in conjunction with the COBE data, we find
\begin{equation}
H_1 = 7.6 \pm 1.4\ .
\label{eq:2}
\end{equation}
Other template choices can modify the constraint slightly but
the errors are  dominated by 
the COBE $7\%$ cosmic
variance errors (\cite{BunWhi97} 1997) and MAXIMA $4\%$ calibration errors on the
temperature fluctuations.

\begin{figure}[htb]
\centerline{\epsfxsize=3.4truein\epsffile{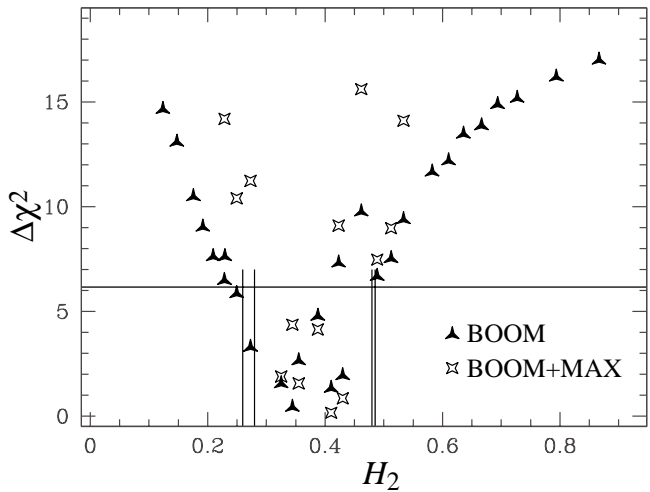}}
\caption{\footnotesize Constraint on the height of the second peak relative to the first: $\Delta \chi^2(H_2)$ 
for the data from $75 < \ell < 600$. 
$1\sigma$ errors are defined as in
Fig.~\protect\ref{fig:l1}.}
\label{fig:h2}
\end{figure}

For $H_2$, we take the data for $75 <\ell < 600$ and consider
templates from models with varying $\Omega_b h^2$, $n$ and 
$\Omega_\Lambda$,
where the last parameter is included to ensure that 
the models reproduce the position of the first peak.   
In this case, $\chi^2$ as a
function of $H_2$ minimized over $\Omega_\Lambda$
exhibits some scatter due to information that is
not contained in the ratio of the peak 
heights (see Fig.~\ref{fig:h2}).  
Nonetheless, the steep dependence of $\Delta \chi^2$ on $H_2$ 
indicates that this statistic is robustly constrained
against the variation of the template.
Taking the outer envelope of $\Delta\chi^2$, we obtain
\begin{eqnarray}
H_2 &=& 0.37 \pm 0.04 \quad ({\rm BOOM}) \,, \nonumber\\
H_2 &=& 0.38 \pm 0.04  \quad ({\rm BOOM+MAX}) \,.
\label{eq:3}
\end{eqnarray}

\begin{figure}[htb]
\centerline{\epsfxsize=3.4truein\epsffile{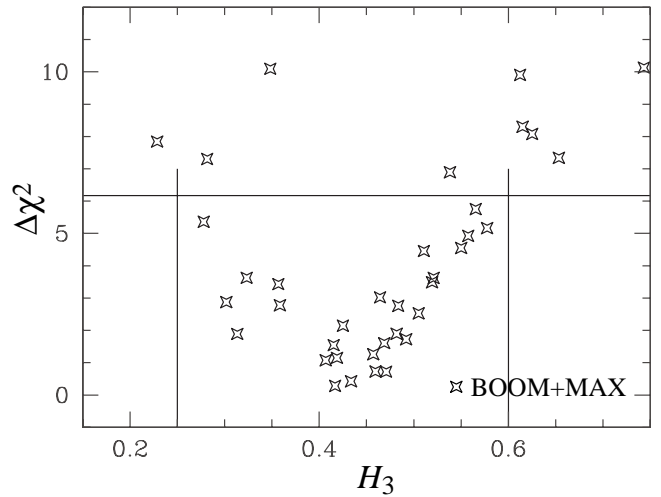}}
\caption{\footnotesize Constraints on the height of the third peak relative to the first: 
$\Delta \chi^2 (H_3)$ for the
data from $75 < \ell < 375$ and $600 < \ell < 800$.
$1\sigma$ errors are defined as in
Fig.~\protect\ref{fig:l1}.}
\label{fig:h3}
\end{figure}

$H_3$ is only weakly constrained by the two highest $\ell$
points ($600 < \ell < 800$) from MAXIMA in conjunction with
the first peak data ($75 < \ell < 375$) from both experiments. 
We consider
templates from models with varying $\Omega_m h^2$, $n$,  
$\Omega_b h^2$ and $\Omega_\Lambda$.  The latter two parameters
are included to ensure that the position of the first peak and the depth of the first trough can be modeled.
Minimizing $\chi^2$ over these two parameters, we plot $\Delta\chi^2$ as
a function of $H_3$ (see Fig.~\ref{fig:h3}) to obtain the bound
\begin{equation}
H_3 = 0.43 \pm 0.07 \quad ({\rm BOOM+MAX}).
\end{equation}
Note that the constraints on $H_3$ employ a template-based extrapolation:
points on the rise to the third peak are used to infer its height. 

\begin{figure}[htb]
\centerline{\epsfxsize=3.4truein\epsffile{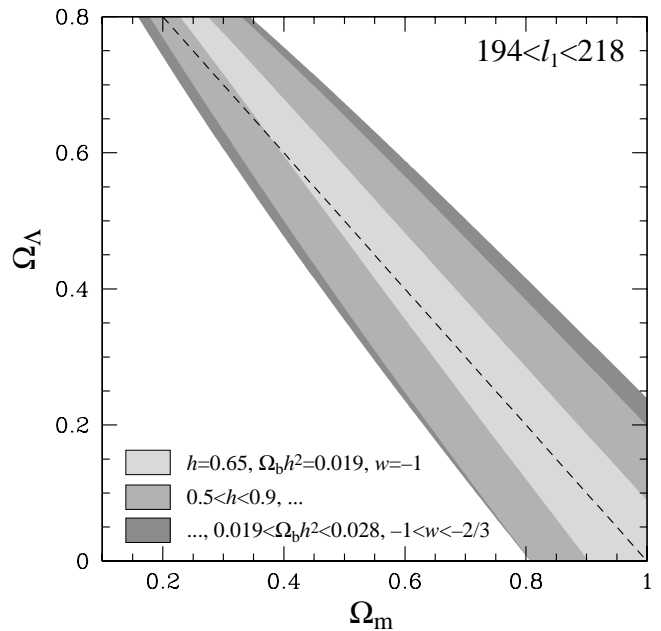}}
\caption{\footnotesize Peak position constraint in the ($\Omega_m$, $\Omega_\Lambda$) plane
with various priors.  The prior assumptions weaken from the light shaded region to the
dark shaded region and '...' means that the unlisted priors are unchanged from the neighboring
region of stronger priors. The constrained region  is strongly limited by the range 
in $h$ considered and to a lesser extent
that in $\Omega_b h^2$ and the equation of state of $\Lambda$, $w$.
} 
\label{fig:l1k}
\end{figure}

\subsection{Cosmological Implications}

The values for the four observables we obtained above can be used 
to derive and understand constraints on cosmological parameters from
the experiments.

The position of the first peak as measured by $\ell_1$ 
is determined by the ratio of the
comoving angular diameter distance to the last scattering epoch 
and the sound horizon at that epoch (\cite{HuSug95} 1995). 
Therefore it is a parameter that depends only on geometry and
sound wave dynamics [see Appendix, eq.~(\ref{eqn:correction})] through 
$\Omega_m+\Omega_\Lambda$,
$h$, $\Omega_m$, $w$, and $\Omega_b h^2$, 
in decreasing order of importance.  The equation of state 
parameter $w=p_\Lambda/\rho_\Lambda$ with $p_\Lambda$ and $\rho_\Lambda$
the pressure and energy density of the vacuum $(w=-1)$ or negative-pressure
energy; we use $\Lambda$ to refer to either option.
The effect of tilt is
small [see eq.~(\ref{eqn:tiltl1})]
\begin{equation}
{\Delta \ell_1 \over \ell_1} \approx 0.17(n-1) 
\end{equation}
and so we neglect it when considering models with $n\sim 1$.

Figure~\ref{fig:l1k} displays the constraints in the $\Omega_m-\Omega_\Lambda$
plane.  Notice that the confidence region is determined
not by uncertainties in the measured value of $\ell_1$, but rather the prior
assumptions about the acceptable range in $h$, $\Omega_b h^2$ and $w$
(\cite{Lanetal00} 2000; \cite{TegZal00b} 2000b).
Given the broad consistency of the data with 
flat models 
($\Omega_\Lambda + \Omega_{m} = 1$)  
with $w = -1$, we will hereafter restrict 
ourselves to this class of models unless otherwise stated.

\begin{figure}[htb]
\centerline{\epsfxsize=3.4truein\epsffile{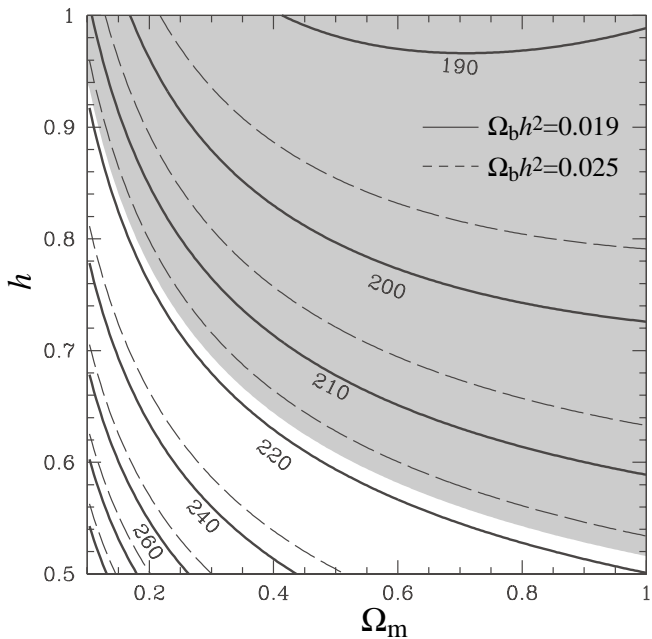}}
\caption{\footnotesize Peak positions in the ($\Omega_m$,$h$) plane for
$\Omega_bh^2=0.019$ and 0.025. Allowed region is indicated by
hatching and assumes $\Omega_b h^2 \ge 0.019$.}
\label{fig:l1omh}
\end{figure}

To better understand the dependence on the Hubble constant, we plot in
Figure~\ref{fig:l1omh} contours of
constant $\ell_1$\ in the ($\Omega_m$,$h$) 
plane for $\Omega_bh^2=0.019$
and $\Omega_bh^2=0.025$. A higher baryon abundance decreases the
sound horizon at last scattering and pushes up
the contours in the direction of higher $\Omega_m$, $h$. 

The 95\% limit, $\ell_1<218$ from 
equation (\ref{eq:1}), excludes the lower left region shown in  
Figure~\ref{fig:l1omh}.
This limit derived from the curve for $\Omega_bh^2=0.019$
is robust in the sense that this baryon abundance
represents the mimimum value allowed by the 
CMB, as we shall see later.
This constraint is summarized approximately by 
$\Omega_m h^{3.8} > 0.079$.  The dependence differs from
the familiar combination $\Omega_{m}h^{2}$ since in a flat universe part of the
effect of lowering $\Omega_{m}$ is compensated for by
the raising of $\Omega_{\Lambda}$.
The lower limit on $\Omega_m h^2$
or equivalently on $\Omega_m$ found in \cite{TegZal00b} (2000b) 
reflects the {\it prior upper} limit on $h$, e.g. for $h<0.8$,
$\Omega_m h^2 > 0.12$ or $\Omega_m > 0.18$.

The 95\% confidence lower limit of the combined fit is $\ell_1>194$,
but it does not give a robust limit in this plane in the absence of
an upper limit on $\Omega_b h^2$.  
The lower limit is also less robust in the sense that it comes
mainly from the MAXIMA result whereas both experiments agree on
the upper limit in the sense that the addition of MAXIMA only 
serves to enhance the confidence at which we may regect larger
values.  

\begin{figure}[htb]
\centerline{\epsfxsize=3.4truein\epsffile{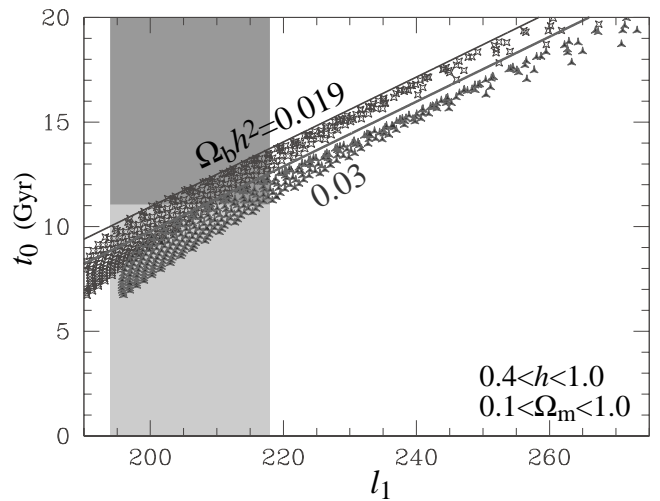}}
\caption{\footnotesize Age-$\ell_1$ correlation.  Points represent models in
the $(\Omega_m, h)$ plane.  Lines represent the simple fit
to the outer envelope given in eqn.~\protect{(\ref{eqn:age})} for the
two choices of $\Omega_b h^2$.
Light shaded region represents current constraints on $\ell_1$;
dark shaded region is also consistent with $t_0 > 11$ Gyr.}
\label{fig:age}
\end{figure}

For a flat geometry, the position of the first peak is 
strongly correlated with the age of the universe.  
The correlation is accidental since $\ell_1$ is
the ratio of the {\it conformal ages} $\int dt/a$ 
at last scattering  and the present that enters, not simply the
physical age today. 
Nonetheless, Fig.~\ref{fig:age} shows that 
the correlation is tight across
the ($\Omega_m$, $h$) values of interest.  The upper envelope corresponds
to the lowest $\Omega_m$ $(=0.1)$ and implies
\begin{eqnarray}
\left({t_0 \over 1 {\rm Gyr}}\right) 
       &\le& 19.7 - 0.155 (250-\ell_1) (0.68)^{1+w} \nonumber\\
&&	- \left( \Omega_b h^2 \over 0.019 \right)^{1.7}\,,
\label{eqn:age}
\end{eqnarray}
where we have included a weak scaling with $w$.
The observations imply $t_0 < 13-14$ Gyr if $\Omega_b h^2 \ge 0.019$.
While this is a weak constraint given the current observational uncertainties,
notice that the central value of the BOOMERanG results $\ell_1\approx 
200$ would imply $t_0 \approx 9-11$Gyrs.  

\begin{figure}[htb]
\centerline{\epsfxsize=3.4truein\epsffile{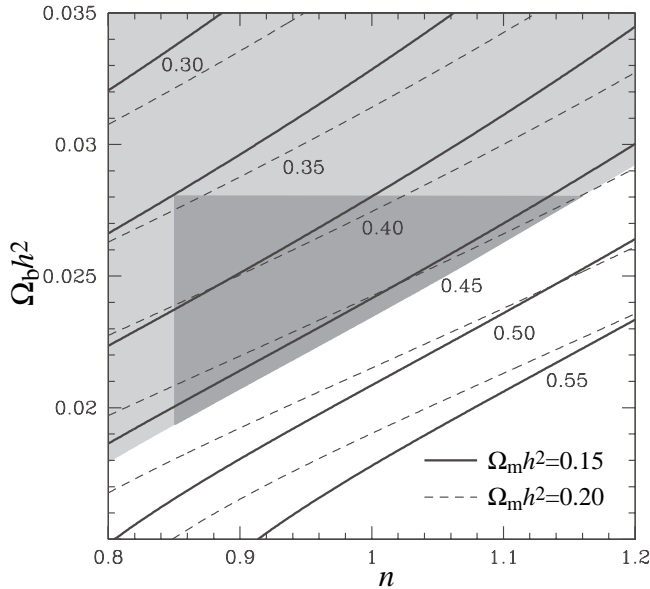}}
\caption{\footnotesize Peak height ratio $H_2$ in the ($n$,$\Omega_b h^2$) plane for
$\Omega_mh^2=0.15$ and 0.20. Allowed region is indicated by light shading and assumes $\Omega_m h^2 \ge 0.15$.  
The constraint
at $n\approx 1$ is nearly independent of this assumption.
Dark shading indicates
the region consistent with $H_1$ ($n>0.85$) and nucleosynthesis
($\Omega_b h^2 < 0.028$).  
}
\label{fig:h2nb}
\end{figure}

The $H_2$ statistic, the ratio of the heights of the
second peak to the first, mainly depends on the
tilt parameter and the baryon abundance. 
This combination is insensitive to
reionization, the presence of tensor modes or any effects that are
confined to the lowest multipoles.
The remaining sensitivity is to $\Omega_mh^2$ and is 
modest in a flat universe due to the cancellation of two
effects [see Appendix eq.~(\ref{eqn:h2form})].  
Figure \ref{fig:h2nb} shows contours of
$H_2=$const. in the ($n,\Omega_b h^2$) plane
for $\Omega_mh^2=0.15$ and $0.20$.  

The result from the previous subsection, $H_2<0.46$ at 
a 95\% confidence level
is shown with shades, giving a constraint
\begin{equation}
\Omega_bh^2> 0.029 \left( {\Omega_m h^2 \over 0.15} \right)^{-0.58}
	(n-1) + 0.024\,.
\label{eq:b-n}
\end{equation}
Note that the constraint at $n=1$ is approximately
independent of $\Omega_m h^2$. 
This limit agrees well with the those from the detailed
likelihood analysis of \cite{TegZal00b} (2000b, c.f. their
Fig. 4) under the same assumptions for $\Omega_m h^2$ and supports
the claim that $H_2$ captures most of the information from the data on
these parameters.
An upper limit from $H_2>0.32$ also exists but is weaker
than conservative constraints from nucleosynthesis 
as we discuss in the next section. 

We can derive a limit on $n$ from the indicator 
$H_1$. 
The cosmological parameter 
dependence of $H_1$
is more complicated than the other two we discussed
above. Fortunately, most complications tend to decrease
$H_1$ by adding large angle anisotropies.  The {\it lower}
limit on $n$ from the lower limit on $H_1$ in the absence
of e.g.\ reionization or tensor modes is therefore conservative.
The upper limit on $n$ is very weak unless one excludes the
possibility of tensor modes as a prior assumption (c.f.
\cite{Baletal00}). 
We search for the minimum $n$ that
gives $H_1$ larger than the 2$\sigma$ lower limit $H_1>4.8$
along the parameter space that maximizes $H_2$. This gives 
a conservative lower bound of $n>0.85$.  This bound
is to good approximation independent of $\Omega_m h^2$ for 
$\Omega_m h^2 \gsim 0.15$. Below this value, the bound
tightens marginally due to the integrated Sachs-Wolfe (ISW) 
effect on COBE scales,  
but in such a way so as 
to maintain the bound 
$\Omega_bh^2>0.019$ when combined with
the constraint from $H_2$, the inequality~(\ref{eq:b-n}).
The analysis of \cite{TegZal00b} (2000b) yields
$n>0.87$ in the same parameter and data space indicating that not 
much information is lost in our much cruder parameterization.
  
\begin{figure}[htb]
\centerline{\epsfxsize=3.4truein\epsffile{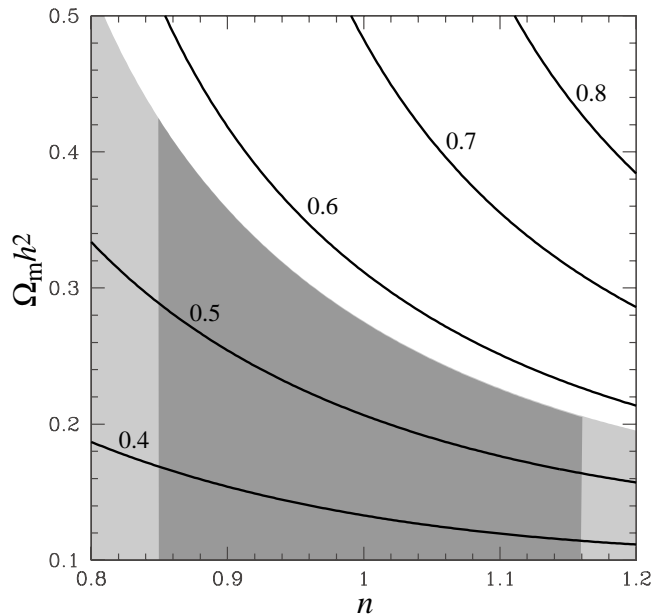}}
\caption{\footnotesize Peak height ratio $H_3$ in the ($n$,$\Omega_m h^2$) plane for
$H_2=0.38$. Allowed region is indicated by light shading. 
Darker shading indicates the allowed region with additional
constraints from $H_1$ $(n>0.85)$ 
and nucleosynthesis $(\Omega_b h^2 < 0.028$, which gives $n<1.16$).}
\label{fig:h3nm}
\end{figure}

The $H_3$ statistic depends more strongly on $\Omega_m h^2$ and $n$
since the baryons affect the height of the third and first peak 
similarly.  In Fig.~\ref{fig:h3nm}, we show the constraint in the
($n$,$\Omega_m h^2$) plane with the baryon density fixed by 
requiring $H_2=0.38$.   When combined with the constraint on
the tilt $n>0.85$, we obtain $\Omega_m h^2 < 0.42$.

In Fig.~\ref{fig:datamodels} (dashed lines), we compare a model
designed to have acceptable values for $\ell_1$, $H_1$, $H_2$
and $H_3$ with the power spectrum data from COBE, BOOMERanG and
MAXIMA.  This gives $\chi^2 = 27.2$ for 30 data points and compares well with
the best fit model of \cite{TegZal00b} (2000b) with their ``inflation prior''
where $\chi^2 = 26.7$.
We summarize the constraints from the CMB as: 
$\Omega_m h^{3.8}$ $>0.079$ (or $t_0<14$ Gyr);
$n>0.85$; $\Omega_b h^2>0.019$; $\Omega_m h^2 < 0.42$.

\section{External Constraints}

In this section, we combine the constraints from the CMB with those from four
other observations: the light element abundances as interpreted by 
BBN theory, the present-day cluster abundance, the
cluster baryon fraction and age of the universe.  
We then translate these constraints onto
the ($\Omega_m$,$h$) plane and discuss consistency checks. 

\subsection{Nucleosynthesis}

The first external constraint we consider is that on baryon abundance
from primordial nucleosynthesis (see \cite{CopSchTur95} 1995; 
\cite{OliSteWal99} 1999; \cite{Tytetal00} 2000 for recent 
reviews).  
\cite{OliSteWal99}\ (1999) give a high baryon density 
option of $0.015\leq \Omega_bh^2\leq 0.023$,
and a low baryon density option $0.004\leq \Omega_bh^2\leq 0.010$ as
a $2\sigma$ range.
A low baryon density is indicated by the traditional 
low value of helium abundance ($Y_p=0.234\pm0.003$) 
(\cite{OliSteWal99} 1999), and agrees with a literal interpretation of
the lithium abundance.
There are also two Lyman limit systems 
which taken at face value point to high deuterium abundance 
(\cite{Sonetal94} 1994; 
 \cite{Caretal94} 1994;
 \cite{Buretal99a} 1999a;
 \cite{Tytetal99} 1999) and implies a low baryon density. 

Our lower limit on $\Omega_b h^2>0.019$ from $H_2$ and $H_1$ 
is strongly inconsistent with the low baryon abundance option.   
In fact our limit is only 
marginally consistent with even the high baryon option 
if we take the latest determination of the deuterium abundance at
face value and treat the individual errors on the systems as 
statistical: D/H$=(3.4\pm0.25)\times 10^{-5}$ from
3 Lyman limit systems (\cite{Kiretal00} 2000) which implies
$\Omega_b h^2 = 0.019 \pm 0.0012$ (\cite{Tytetal00} 2000; 
\cite{Buretal99b} 1999b).
In this paper, we provisionally accept the 2$\sigma$ limit of
\cite{OliSteWal99} (1999).  The issue, however, is clearly
a matter of systematic errors, and we discuss in what follows 
where they could appear, trying to find a conservative upper limit.

Since it is possible that the measured D/H abundance is high
due to contamination by $H$, we consider the firm lower limit on
the D/H abundance from interstellar clouds.
The earlier UV data (\cite{McCetal92} 1992) show a variation of
D/H from 1.2 to $2.5\times10^{-5}$. This variation is confirmed
by modern high resolution spectrographs. The clouds studied are
still few in number and range from 
D/H=(1.5$\pm0.1) \times 10^{-5}$ (\cite{Linsky98} 1998; 
\cite{Linetal95} 1995) to 
$0.7 \times 10^{-5}$ (\cite{Jenetal99} 1999). This variation is reasonable since the clouds
are contaminated by heavy elements, indicating 
significant astration effects. 
Therefore, we take the upper
value as the observational D/H abundance, and take the
minimum astration effect (factor 1.5) from model calculations 
(see \cite{Tos96} 1996; \cite{OliSteWal99} 1999, modified for a
10 Gyr disk age) to infer the lower limit on
the primordial deuterium abundance.  We take $2\times 10^{-5}$ as
a conservative
lower limit on D/H. This value agrees with the pre-solar 
system deuterium abundance inferred from $^3$He (\cite{GeiGlo98}
1998).
This D/H corresponds to $\Omega_bh^2=0.028$ and $Y_p=0.250-0.252$.

The helium abundance $Y_p$ directly depends on
the theoretical calculation of the helium recombination line, and the
discrepancy between the estimate ($Y_p= 0.244 \pm 0.002$,
\cite{IzoThu98} 1998) and  
the traditional estimates ($Y_p=0.234\pm 0.003$) largely arises from
the two different calculations, \cite{Smi96} (1996) and 
\cite{Blo72}
(1972). The helium abundances derived from three recombination lines
He I $\lambda$4471, $\lambda$5876 and $\lambda$6678 for a given HII
region differ fractionally by a few percent.  Also,
the effect of underlying stellar absorption by hot stars is unclear: 
\cite{IzoThu98} (1998) use the departure of the He I $\lambda6678/\lambda5876$ 
strengths from the Smits calculation as an estimator, 
but a calculation is not
actually available for the He line absorption 
effect.   While these variations are usually included as
random errors in the nucleosynthesis literature, we suspect 
that the error in the helium abundance is dominated by
systematics and a further change by a few percent in excess of the
quoted range is not excluded. 

The interpretation of the Li abundance rests on a 
simplistic model of stars.
It seems that our understanding of 
the $^7$Li abundance evolution is still far from 
complete: for instance we
do not understand the temperature gradient of the Li/H ratio in
halo dwarfs, which shows a trend opposite to what is expected 
with $^7$Li destruction due to diffusion. % what is (\cite{Ryaetal66} 1966). 
Hence we do not view the primordial $^7$Li abundance 
determinations as rock-solid.

We therefore consider two cases $\Omega_b h^2 < 0.023$ as a widely
accepted upper limit and $\Omega_bh^2<0.028$ as a very conservative
upper limit based on interstellar deuterium.
When combined with the limit of eq.~(\ref{eq:b-n}), the latter  
constraint becomes $n<1.16$ for $\Omega_m h^2 < 0.2$ (as appropriate
for setting a lower bound on $\Omega_m$ in the next section, 
see also Fig.~\ref{fig:h3nm});
if we instead take $\Omega_bh^2<0.023$ (\cite{OliSteWal99} 1999), 
the limit becomes $n<0.98$. 
In conjunction with the constraint
from $H_{1}$, the allowed range for the tilt becomes
\begin{equation}
0.85<n<1.16 \,.
\end{equation}

\subsection{Cluster Abundance}

The next external constraint we consider is the abundance of clusters
of galaxies, which constrains the matter power spectrum
at intermediate scales.  
%There are a number of other ways to infer the fluctuation
%power at low redshifts, but most of them are subject to
%uncertainties from unknown biasing factors for galaxies;
%or otherwise, they are susceptible to errors of distance indicators. 
%The fluctuation power inferred from the cluster mass or
%temperature function 
%seems to be more robust and seems difficult to avoid in so far as
%we take the hierarchical clustering scenario in the CDM model with
%Gaussian initial conditions. 
We adopt
the empirical fit of \cite{Ekeetal96} (1996) for a flat
universe $\sigma_8=(0.52\pm0.08)\Omega_m^{-0.52+0.13\Omega_m}$. 
The value of $\sigma_8$ is well converged within 1 $\sigma$ among
different authors (\cite{ViaLid99} 1999; \cite{Pen98} 1998). 
This is because the cluster abundance depends strongly
on $\sigma_8$ due to its appearance in the exponential of a
Gaussian in the Press-Shechter formalism.

We take the amplitude at COBE scales with a 14
\% normalization uncertainty (95 \% confidence) together with the 95
\% confidence range coming from the cluster abundance to obtain an allowed
region that is a function of $\Omega_m$, $h$ and $n$ and can be roughly
described by 
\begin{equation}
0.27 < \Omega_m^{0.76} h n < 0.35 \,,
\label{eqn:clusterabundance}
\end{equation}
assuming no tensor contribution to COBE and $\Omega_b h^2 = 0.028$. 
These assumptions lead to the most conservative
constraints on the ($\Omega_m$,$h$) plane. The lower limit
comes from undershooting $\sigma_8$ which is only exacerbated
with the inclusion of tensors.  It also depends on the upper limit on 
$n$, which is maximized
at the highest acceptable baryon density $\Omega_b h^2=0.028$. 
The upper limit comes from overshooting $\sigma_8$
and depends on the lower limit on $n$, which only tightens 
with the inclusion of tensors and lowering of the baryon density.  

\subsection{Baryon Fraction}

The third external constraint we consider is 
the baryon fraction
in rich clusters, derived from X-ray observations.
The observed baryon fraction shows a slight
increase outwards, and the true baryon fraction inferred for the
entire cluster depends on the extrapolation. The estimates range
from $(0.052\pm0.0025)h^{-3/2}$ (\cite{WhiFab95} 1995; lowest estimate) to
$(0.076\pm0.008)h^{-3/2}$ (\cite{ArnEvr99} 1999; highest estimate)
for rich clusters.
We take the 2$\sigma$ limits to correspond to these two extreme values.
We remark that very similar constraints are derived from the Sunyaev
Zeldovich effect for clusters as long as $h=0.5-1.0$: 
\cite{Myeetal97} (1997) derive 
$(0.061\pm0.011)h^{-1}$, and \cite{Greetal00} (2000)
give $(0.074\pm0.009)h^{-1}$. Adding baryons locked into stars to that
in gas inferred by $X$-ray observations, and 
assuming the cluster baryon fraction represents the global
value (\cite{Whietal93} 1993), we have $f_b\equiv \Omega_b/\Omega_m$
constrained as 
\begin{equation}
 0.052 h^{-3/2} + 0.006 h^{-1} < f_b
<
 0.076 h^{-3/2} + 0.015 h^{-1}
\label{eqn:baryonfraction}
\end{equation}
This relation is used to convert the constraints on $\Omega_bh^2$
into the $\Omega_m$ vs $h$ plane.

\subsection{Age}

We take the lower limit on the age of the universe
to be $t_0 > 11$Gyr based on stellar evolution.
While this is not based on statistical analysis, no authors have 
ever claimed a cosmic age less than this value
(\cite{Graetal97} 1997; \cite{Rei97} 1997; \cite{Chaetal98} 1998).

\subsection{Allowed Region}

\begin{figure}[htb]
\centerline{\epsfxsize=3.4truein\epsffile{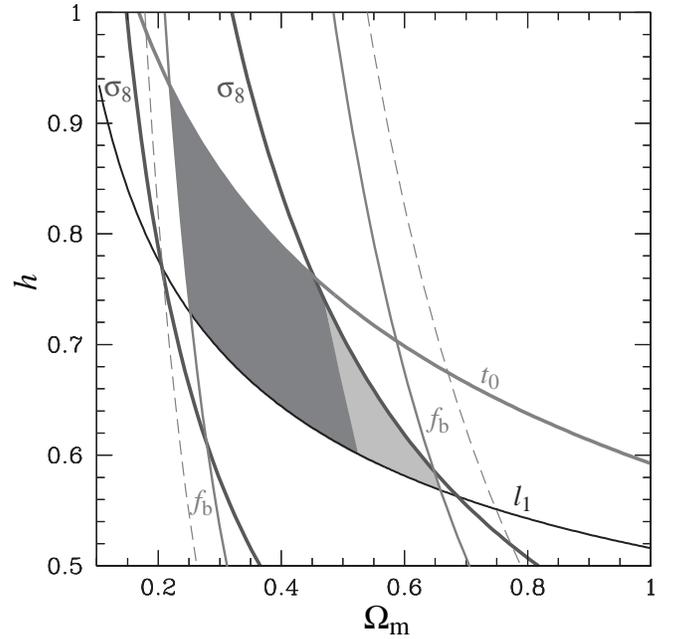}}
\caption{\footnotesize Summary of constraints. The shaded region is allowed
by all constraints considered in this paper. The dark shaded
region contains models that are also consistent with 
$\Omega_b h^2 < 0.023$ (see text).}
\label{fig:sum}
\end{figure}

We display all our constraints in the ($\Omega_m$,$h$) plane
in Fig. \ref{fig:sum}. 

Combining the range $0.019 \le \Omega_bh^2 \le 0.028$ from BBN and 
the CMB,
together with the constraint on $\Omega_b/\Omega_m$ from the
baryon fraction (\ref{eqn:baryonfraction})
leads to the range 
\begin{equation}
{0.019 \over 0.076h^{1/2}+0.015h}< \Omega_m <
{0.028 \over 0.052h^{1/2}+0.006h}\,,
\end{equation}
which is plotted as the solid contours labeled $f_b$ in Fig.~\ref{fig:sum}.
We also plot the more conservative limits derived from taking 
the $2\sigma$ extremes of the extreme baryond fraction measures 
($0.076 \rightarrow 0.092$ and $0.052 \rightarrow 0.047$) as dashed
lines.

We convert the cluster abundance constraint
using the range in tilts acceptable from the CMB constraints
and the limit $\Omega_b h^2 < 0.028$ from BBN ($0.85 < n < 1.16$)
and find 
\begin{equation}
0.15 < \Omega_m h^{1.3} < 0.32 \,.
\end{equation}
This range is displayed by the contours labelled ``$\sigma_8$''.
Finally the constraints $t_0 > 11$Gyr and $\ell_1 < 218$ are labelled
as ``$t_0$'' and ``$\ell_1$'' respectively.

The shading indicates the parameter space within which a
model consistent with the CMB and external constraints may be
constructed.  Dark shading indicates the region that is also
consistent with the stronger nucleosynthesis bound of 
$\Omega_b h^2 < 0.023$.  This does not mean that all models in this
region are consistent with the CMB data.  To construct a viable model for a given
($\Omega_m$,$h$) in this region, 
one picks a tilt $n$ between $0.85-1.16$ consistent with the cluster abundance
constraint (\ref{eqn:clusterabundance}) and then a baryon 
density consistent with $H_2$ (\ref{eq:b-n}) and $\Omega_b h^2 < 0.028$
(or $0.023$).
In Fig.~\ref{fig:datamodels} (solid lines), we verify that the 
power spectrum prediction of a model so constructed is a good fit to the data. 
Here $\chi^2 = 28.5$ for the 30 data points to be compared with 
$\chi^2 =27.2$ for the model optimized for the CMB alone.

\subsection{Consistency Checks}

There are a variety of other cosmological measurements that
provide alternate paths to constraints in the ($\Omega_m$,$h$) plane. 
We do not use these measurements as
constraints since a proper error analysis requires 
a detailed consideration of systematic errors 
that is beyond the scope of this paper.  
We instead use them as consistency checks on 
the adiabatic CDM framework.

{\it Hubble constant.}
A combined analysis of secondary distance indicators gives 
$h=0.71\pm0.04$ for an assumed LMC distance of 
50 kpc (\cite{Mouetal99} 2000).
Allowing for a generous 
uncertainty in the distance to the LMC 
(see \cite{Fuk00} 2000 for a review) these values may be
multiplied by $0.95-1.15$ and this should be compared with our
constraint of $0.6 < h < 0.9$.  

{\it Cosmic acceleration.}
The luminosity distance to distant supernovae requires
$\Omega_m<0.48$ for flat $\Lambda$ models if the systematic errors are no
worse than they are claimed (\cite{Rieetal98} 1998; \cite{Peretal99} 1999). 
This limit should be
compared with our constraint of $\Omega_m < 0.6$.

{\it  Mass-to-Light Ratio.} The $\Omega_m$ constraint
 we derived using 
the range $0.019 < \Omega_b h^2 < 0.023$ (see Fig.~\ref{fig:sum}, 
dark shaded region) and the cluster
baryon fraction corresponds 
to $M/L_B=(350-600)h$, which
is roughly consistent with $M/L_B$ for rich 
clusters (e.g. \cite{CarYeeEll97} 1997). A yet larger $\Omega_m$ 
($\Omega_m>0.45$) would
imply the presence of a substantial
amount of matter outside clusters and galaxies, whereas
we have some evidence indicating the contrary (\cite{Kaietal98} 1998). 

{\it  Power Spectrum.}
The shape parameter of the transfer function is $\Gamma\approx
\Omega_mh\exp[-\Omega_b(1-1/\Omega_m)]\approx 
0.22-0.33$ for our allowed region (\cite{Sug95} 1995).  
This is close to the value that fits the galaxy
power spectrum;
$\Gamma=0.2-0.25$ (\cite{Efsetal90} 1990; \cite{PeaDod94} 1994).
On smaller scales, the Ly$\alpha$ forest places constraints on
the amplitude and slope of the power spectrum near $k \sim 1$ $h$
Mpc$^{-1}$
at $z \sim 3$ (\cite{Croetal99} 1999; \cite{McDetal99} 1999).   
\cite{McDetal99} (1999) map these constraints onto cosmological parameters 
within $\Lambda$CDM as $n = 0.93 \pm 0.10$ and
$\sigma_8 = 0.68 + 1.16(0.95-n) \pm 0.04$.

{\it  Cluster Abundance Evolution.}
The matter density $\Omega_m$
can be inferred from evolution of the rich cluster abundance
(\cite{OukBla92} 1992), but the result depends sensitively 
on the estimates of
the cluster masses at high redshift. Bahcall \& Fan (1998) 
argue for a low
density universe $\Omega_m=0.2{+0.3 \atop -0.1}$; 
\cite{BlaBar98} (1998)
and \cite{Reietal99} (1999) favor 
a high density $\Omega\approx 1$, while Eke et al. (1998)
obtain a modestly low density universe $\Omega_m=0.36\pm0.25$.

{\it Peculiar Velocities.}
The results from peculiar velocity
flow studies are controversial: they vary from $\Omega_m=0.15$ to 1
depending on scale, method of analysis and
the biasing factor (see e.g. \cite{Dek99} 1999 for a recent review).

{\it Local Baryons.}
The CMB experiments require a high baryon abundance. The lower
limit (together with a modest red-tilt of the spectrum)
is just barely consistent with the high baryon abundance option from 
nucleosynthesis.  
The required baryon abundance is still below the maximum
estimate of the baryon budget in the local universe
$0.029h^{-1}$ (\cite{FukHogPee98} 1998), but this requires 3/4 of the baryons
to reside near groups of galaxies as warm and cool gas.

\section{Future Directions and Implicit Assumptions}

A useful aspect of our approach is that one can ask 
how the allowed parameter space might evolve as the data evolves.  
More specifically, what aspect of the data can make the allowed
region qualitatively change or vanish altogether?  If the data
are taken at face value, what theoretical assumptions might be modified 
should that come to pass?

An increase in the precision with which the acoustic scale
is measured may lead to a new age crisis.  It is noteworthy
that the secondary peaks will eventually provide a substantially
more precise determination of the scale due to sample 
variance limitations per patch of sky, the multiplicity of peaks, and
the effects of driving forces and tilt on the first peak [see Appendix
eq.~(\ref{eqn:harmonicseries})].  
Indeed, consistency between the determinations of this scale
from the various peaks will provide a strong consistency check
on the underlying framework.  If the measurements were to
determine an equivalent $\ell_1 \le 200$, then $t_0 < 10-11$Gyrs in 
a flat $\Lambda$ cosmology with $\Omega_b h^2=0.019$; taking
$\Omega_b h^2 =0.03$ {\it decreases} the age by 1 Gyr and exacerbates
the problem.  Such a crisis, should it occur, can only be mildly 
ameliorated by replacing the cosmological constant with a
dynamical ``quintessence'' field.  Because increasing the equation
of state $w$ from $-1$ reduces both $\ell_1$ and the age, only
a relatively extreme choice of $w \gsim -1/3$ can help substantially
[see eqn.~(\ref{eqn:age})].  This option would also imply that
the universe is not accelerating and is in conflict with evidence from
distant supernovae.  However, other solutions may be 
even more unpalatable: a small positive curvature 
{\it and} a cosmological constant or a delay in recombination.  

As constraints on the tilt improve by extending the dynamic range 
of the CMB observations and those on $H_2$ by resolving the second 
peak, one might be faced with a baryon crisis.  Already $\Omega_b h^2=0.019$
is only barely allowed at the 95\% CL.  Modifications of 
big-bang nucleosynthesis that allow a higher baryon density for the
same deuterium abundance are difficult to arrange: current directions of
study include inhomogeneous nucleosynthesis (e.g. \cite{KaiKurSih99} 1999)
and lepton asymmetry (e.g. \cite{LesPel00} 2000; \cite{Espetal00} 2000).  
On the CMB side, there are two general 
alternatives.  The first possibility is that there is a smooth
component that boosts the relative height of the first peak
(\cite{Bouetal00} 2000).  That possibility can be constrained in the
same way as tilt: by extending the dynamic range, one can distinguish
between smooth and modulated effects.  The direct observable in 
the modulation is the ratio of energy densities in non-relativistic
matter that is coupled to the CMB versus the CMB itself [$R_*$, 
see eqn.~(\ref{eqn:matterbaryon})],
times the gravitational potential, all evaluated at last scattering.  
The second possibility is that one of the links in 
the chain of reasoning from the observables to the 
baryon and matter densities today is broken in some way.

It is noteworthy that there is no aspect of the CMB data today
that strongly indicates missing energy in the form of a cosmological
constant or quintessence.  An Einstein-de 
Sitter universe with a high baryon density is still viable unless external
constraints are introduced.  Under the assumption of a flat $\Lambda$ cosmology, 
tight constraints on $\Omega_{m} h^{3.8}$ from the peak locations
and $\Omega_{m}h^{2}$ from the third and higher peak heights should allow
$\Omega_m$ and $h$ to be separately measured. It will be important to check 
whether the CMB implications for $\Omega_m$ are consistent with external 
constraints.
 
Aside from acceleration measurements from distant supernovae, the missing
energy conclusion finds its strongest support from the 
cluster abundance today through $\sigma_8$ and the cluster
baryon fraction. 
Changes in the interpretation of these measurements would affect the viability of
the Einstein-de Sitter option. 

The interpretation of the cluster abundance is based on the assumption
of Gaussian initial conditions and the ability to link the power
spectrum today to that of the CMB through the usual transfer
functions and growth rates.  One possibility is that the primordial
power spectrum has strong deviations from power-law behavior (e.g. \cite{AdaRosSar97} 1997).
Just like tilt, this possibility can be constrained through the
higher peaks.  

A more subtle modification would arise if the neutrinos had a 
mass in the eV range.  Massive
neutrinos have little effect on the CMB itself 
(\cite{DodGatSte96} 1996; \cite{MaBer95} 1995) but strongly
suppress large scale structure through growth rates
(\cite{Jinetal93} 1993; \cite{Klyetal93}  1993).  
A total mass (summed over neutrino species) 
of $\sum {m_{\nu_i}}=1$eV would be sufficient to 
allow an Einstein-de Sitter universe in
the cluster abundance.  One still violates the cluster
baryon fraction constraint.  In fact, even for lower $\Omega_m$ one
can only find models consistent with both the cluster
abundance and baryon fraction if $\sum {m_{\nu_i}}<4$eV. 
These constraints could be weakened if some unknown form of support 
causes an underestimate of the dark mass in clusters through
the assumption of hydrostatic equilibrium.  
They could also be evaded if modifications in nucleosynthesis
weaken the upper limit on the baryons.

\section{Conclusions}

We find that the current status of CMB power spectrum
measurements and their implications for cosmological
parameters can be adequately summarized with four numbers:
the location of the first peak $\ell_1=206\pm 6$ and the
relative heights of the first three peaks $H_1=7.6 \pm 1.4$, $H_2=0.38\pm 0.04$
and $H_3=0.43\pm 0.07$.  
When translated into cosmological parameters, they imply
$\Omega_m h^{3.8} > 0.079$ (or $t_0<13-14$ Gyr),
$n>0.85$, $\Omega_b h^2>0.019$, $\Omega_m h^2 < 0.42$ for
flat $\Lambda$CDM models.  
   Other constraints mainly reflect
the implicit (with priors) or explicit use of information
from other aspects of cosmology. 
For example, our consideration of nucleosynthesis, the
cluster abundance, the cluster baryon fraction, and the age
of the universe leads to an allowed region where
$0.6 < h <0.9$, 
$0.25 < \Omega_m < 0.45$, 
$0.85 < n < 0.98$,
$0.019 < \Omega_b h^2 < 0.023$.   The region is narrow, but there
clearly are adiabatic CDM models viable at the $95\%$ CL as exemplified
in Fig.~\ref{fig:datamodels}.  The region widens and the quality of the
fit improves if one allows somewhat higher baryons 
$\Omega_b h^2 < 0.028$ as discussed in this paper.  With this 
extension the tilt can be larger than unity $n<1.16$ and $\Omega_m$ as high
as 0.6.
We note that in both cases our limits reflect conservative
assumptions about tensors and reionization, specifically that they are
negligible effects in the CMB.\footnote{This assumption is
{\it not} conservative when considering likelihood constraints from the CMB 
alone.  The presence of tensors substantially weaken the upper limit on $n$.}

The constraints on these and other CMB observables are expected to
rapidly improve as new data are taken and analyzed.  We have identified sets
of observables that should provide sharp consistency tests 
for the assumptions that underly their translation into 
cosmological parameters in the adiabatic CDM framework.  

With the arrival of precision data sets, the enterprise of 
measuring cosmological parameters from the CMB has entered a new era.
Whether the tension between the observations that is confining the standard
parameters to an ever tightening region is indicating convergence to a final
solution or hinting at discord that will challenge our underlying assumptions
remains to be seen. 

\acknowledgements

We would like to thank M. Hudson, S. Landau, 
G. Steigman, M. Turner, and S. Weinberg for 
useful discussions.
WH is supported by the Keck Foundation;  MF 
by the Raymond and Beverly Sackler Fellowship in Princeton; MZ 
by the Hubble Fellowship 
HF-01116-01-98A from STScI, operated by
AURA, Inc. under NASA contract NAS5-26555; MT by
NASA grant NAG5-9194 and NSF grant AST00-71213.

\appendix

\section{Scaling Relations}

The phenomenology of the peaks can be understood through 
three fundamental scales which vary with cosmological parameters:
%in analytically calculable ways:
the acoustic scale $\ell_A$, the equality scale $\ell_{\rm eq}$
and the damping scale $\ell_D$.  

We begin by employing an idealized picture of the photon-baryon
fluid before recombination that neglects dissipation and time variation of 
both the sound speed $c_s$ and the gravitational driving forces. 
Simple acoustic physics then tells us that the
effective temperature perturbation in the wavemode $k$ 
oscillates as  (\cite{HuSug95} 1995)
\begin{equation}
\Delta T(\eta_*,k) = [\Delta T(0,k) + R_{*}\Psi]\cos(ks_*) 
	- R_{*}\Psi\,.
\label{eqn:oscillator}
\end{equation}
where the sound horizon at the last scattering surface 
$s\equiv \int c_{s}d\eta = \int c_s dt/a$ with
$c_{s}^{2} = 1/3(1+R)$ and $R = 3\rho_{b}/4\rho_{\gamma}$.
$\Psi$ is the gravitational potential.
The asterisk denotes evaluation at last scattering.
Baryons modulate the amplitude of the oscillation by shifting
the zero point by $R_{*} \Psi$.
The result is that the modes that reach maximal compression
inside potential wells at last scattering 
are enhanced over those that reach
maximal rarefaction.  Note that this amplitude modulation 
is not equivalent to saying that the hot spots are enhanced 
over cold spots as  the same reasoning applies to potential ``hills''.

The oscillator equation (\ref{eqn:oscillator}) 
predicts peaks in the angular power spectrum at
$\ell_m = m \ell_A$ where
$\ell_A$ is related to $s_*$ through its projection on the sky today
via the comoving angular diameter distance (\cite{HuSug95} 1995)
\begin{eqnarray}
    D &\approx & 2
    {[1+\ln(1-\Omega_\Lambda)^{0.085}]^{1+1.14(1+w)}
	\over 
    \sqrt{\Omega_{m}H_{0}^{2} \Omegat^{(1-\Omega_\Lambda)^{-0.76}}}}
\nonumber\\
      &\equiv&
        { 2d \over \sqrt{\Omega_{m} H_{0}^2}} 		 \,,
\end{eqnarray}    
where $\Omega_\Lambda$ refers to the density in dark energy
with a fixed equation of state $w = p_\Lambda/\rho_\Lambda$ ($w=-1$ for a 
true cosmological constant) and the total density is $\Omegat = 
\Omega_{m} + \Omega_\Lambda$.  For convenience, we have defined
the dimensionless angular diameter distance $d$ which 
scales out the effect of the expansion rate during 
matter domination; hence it is equal to unity for an 
Einstein-de Sitter cosmology. 
More specifically, $\ell_A \equiv \pi D/s_*$ or
\begin{eqnarray}
\label{eqn:correction}
    \ell_{A} &\approx&
        {172 d} \left({z_* \over 10^3}\right)^{1/2}  \\
%\nonumber
%        ,\\
&& \times 
\left( {1 \over \sqrt{R_*}} \ln
        {\sqrt{1+R_*} +  \sqrt{R_* + r_* R_*}
        \over  1 + \sqrt{r_* R_*}} \right)^{-1}  \,,
\nonumber
\end{eqnarray}
where the radiation-to-matter and baryon-to-photon ratios
at last scattering are
\begin{eqnarray}
r_*     &\equiv& \rho_{r}(z_{*})/\rho_{m}(z_{*})  
               =  0.042 \wm^{-1} (z_*/10^3)\,, \nonumber\\
R_*     &\equiv& 3\rho_{b}(z_{*})/4\rho_{\gamma}(z_{*})
               = 30 \wb (z_*/10^3)^{-1}\,
\label{eqn:matterbaryon}
\end{eqnarray}
with a redshift of last scattering given by 
\begin{eqnarray}
z_* &\approx& 1008 (1+0.00124 \wb^{-0.74})(1+c_1 \wm^{c_2})\,,
	\nonumber\\
c_1    &=& 0.0783 \wb^{-0.24} (1+39.5 \wb^{0.76})^{-1}\,,\nonumber\\
c_2    &=& 0.56 (1+21.1 \wb^{1.8})^{-1} \,.
\end{eqnarray}
Here we use the shorthand convention $\omega_b=\Omega_b h^2$ and
$\omega_m=\Omega_mh^2$.
Baryon drag works to enhance $m=$odd over $m=$even 
peaks in the power.   

%Here and throughout 
%$\wb\equiv\Omega_b h^2$ and $\wm \equiv \Omega_m h^2$. 	

These simple relations are modified by driving and dissipative
effects.  The driving effect comes from the decay of the gravitational
potential in the radiation dominated epoch which enhances
the oscillations and leads to an increase in power of 
approximately a factor of 20 
for $\ell > \ell_{\rm eq}$  (\cite{HuSug95} 1995)
where
\begin{equation}
    \ell_{\rm eq} \equiv (
    2 \Omega_{m} H_{0}^{2} z_{\rm eq})^{1/2} D
    \approx {438 d} \wm^{1/2} \,.
\end{equation}
It also introduces a phase shift to the oscillations such that
the $m$th peak of a scale invariant ($n=1$ model) is at\footnote{The coefficients are from fits to
the first peak at $\Omega_b h^2=0.02$.  For better accuracy,
replace the coefficient $0.267$ with $0.24$ for $\ell_2$
or $0.35$ for $\ell_3$.  Note that the {\it fractional} change made
by the phase shift decreases with $m$.}
\begin{eqnarray}
    \ell_m &=&
	\ell_A ( m - \phi) \nonumber\\
	\phi &\approx & 
	0.267 \left( {r_* \over 0.3}\right)^{0.1} \,.
\label{eqn:harmonicseries}
\end{eqnarray}
Tilt also mildly affects the location of the peaks especially the
first which is broadened by radiation effects; around
$n=1$ (and $\Omega_m h^2=0.15$), the change is approximately
\begin{eqnarray}
    {\Delta \ell_1 \over \ell_1} &\approx& 0.17 (n-1) \,,\nonumber\\
    {\Delta \ell_2 \over \ell_2} &\approx& 0.033 (n-1)\,, \nonumber\\
    {\Delta \ell_3 \over \ell_3} &\approx& 0.012 (n-1) .
\label{eqn:tiltl1}
\end{eqnarray}
The matter dependence is weak: for $\Omega_m h^2=0.25$ the
coefficient $0.17$ is reduced to $0.15$ for $\ell_1$.

The other effect of radiation driving is to
reduce the baryon drag effect by reducing
the depth of the potential wells at
$z_*$.  The baryon drag effect is fractionally
of order $R_{*}\Psi(\eta_{*})/\Psi(\eta_{\rm initial}) 
\approx R_{*} T(k)$ where $T(k)$ is the matter transfer
function and $k$ is the comoving wavenumber in Mpc$^{-1}$.   
The transfer function 
quantifies the decay of the potential in the radiation
dominated epoch (see e.g. Eisenstein \& Hu 1999 for a fit). 
The break in the transfer function is
also given by the horizon scale at matter-radiation equality
so that it appears on the sky at $\ell_{\rm eq}$.  Shifting
the equality scale to raise $\ell_{\rm eq}$ by raising the matter content
decreases the overall amplitude of the oscillations but increases
the odd-even modulation leading to somewhat counterbalancing
effects on the peak heights.

\begin{figure}[htb]
\centerline{\epsfxsize=3.4truein\epsffile{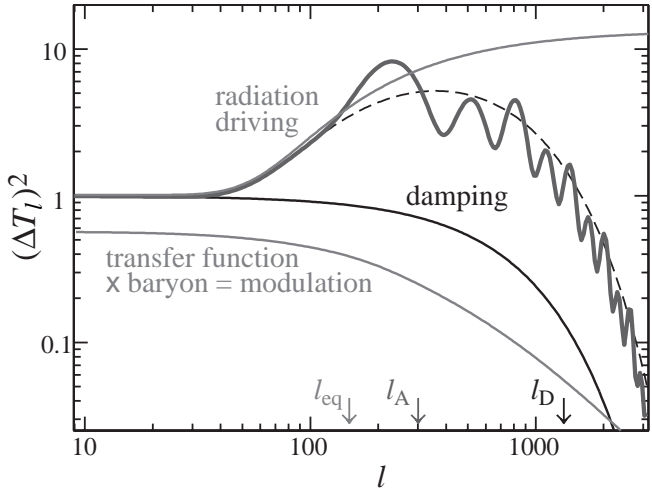}}
\caption{\footnotesize A model power spectrum based on the
fundamental scales $\ell_{\rm A}$, $\ell_{\rm eq}$, 
$\ell_{\rm D}$ and the baryon-photon ratio $R_*$ which modulates
the amplitude of the oscillations as $R_* T(\ell/D)$ where $T(k)$ is
the transfer function.}
\label{fig:mock}
\end{figure}

Finally, the acoustic oscillations are dissipated on small scales.
The quantitative understanding of the effect requires numerical 
calculation but its main 
features can be understood
through qualitative arguments. Since the oscillations dissipate
by the random walk of the photons in the baryons, the characteristic
scale for the exponential damping of the amplitude 
is the geometric mean between the mean free path
$\lambda_C = (x_{e} n_{e} \sigma_{T} a)^{-1}$ and the horizon scale 
\begin{equation}
\eta_* = 2(\Omega_m H_0^2)^{-1/2} z_*^{1/2} 
[\sqrt{1+r_*} -\sqrt{r_*}]\,.
\end{equation}
under the Saha approximation $x_{e} \propto \wb^{-1/2}$
so that $k_D \sim (\eta_* \lambda_C)^{-1/2}
\propto z_*^{5/4} \wb^{1/4}\wm^{1/4}$.  Numerically, the
scaling is slightly modified to (refitting values from 
Hu \& White 1997)
\begin{equation}
\ell_D \equiv k_{D} D\approx { 2240 d
	  \over [(1+r_{*})^{1/2} - r_{*}^{1/2}]^{1/2}
	} \left( { z_* \over 1000} \right)^{5/4}
	\wb^{0.24} \wm^{-0.11}\,.
\end{equation}
Compared with the acoustic scale $\ell_A$, it has a much stronger
dependence on $\wb$ and the redshift of recombination $z_*$.
We show a model spectrum\footnote{The model spectrum is obtained
by following a construction based on 
\cite{HuWhi97} (1997):
the damping envelope is 
\begin{equation}
{\cal D}_{\ell}= \exp[-(\ell/\ell_D)^{1.2}] \,, 
\end{equation}
yielding acoustic oscillations of the form
\begin{equation}
{\cal D}_\ell {\cal A}_\ell =
    [1+R_{*}T(\ell/D)]{\cal D}_\ell \cos[\pi (\ell/\ell_A+\phi)]
     -R_{*}T(\ell/D)\,;
\label{eqn:oscillationform}
\end{equation}
the potential driving envelope is
\begin{equation}
{\cal P}_\ell \approx 1+ 19\exp(-1.4\ell_{\rm eq}/\ell) \,.
\end{equation}
The spectrum is then constructed as
\begin{eqnarray}
    (\Delta T_\ell)^2 &\propto & \left( {\ell \over
    10}\right)^{n-1}
    {\cal P}_{\ell} {\cal D}_{\ell}^{2}
    {1 \over 2}
    \left( { {\cal A}_{\ell}^2 -1 \over 1 + (\ell_A/2\ell)^6} + 2 \right) \,,
\end{eqnarray}
where  we have added an offset to the oscillations to roughly
account for projection smoothing and the Doppler effect and forced
the form to return to ${\cal P}_{\ell}$ above the first
peak to account for the early ISW effect from the
radiation (\cite{HuSug95} 1995).  This mock spectrum should
only be used to understand the qualitative behavior of the
spectrum.  } 
obtained this way for 
a $\Lambda$CDM cosmology with $\Omega_{m}=0.35$,
$\Omegat=1$, $h=0.65$, $\wb=0.02$, and $w=-1$ in Fig.~\ref{fig:mock}.
For this cosmology, the three fundamental scales are
$\ell_{\rm eq}=149$ ($\ell_1=221$), $\ell_{A} = 301$, and
$\ell_{D}=1332$.   The dependence of the morphology of the
acoustic peaks on cosmological parameters is controlled  by
these three scales.  Around
the fiducial $\Lambda$CDM model with the parameters given above
\begin{eqnarray}
    {\Delta \ell_{A} \over \ell_{A}}
    &\approx&
    -0.11 \Delta w 
    -0.24 {\Delta \wm \over \wm}
    +0.07 {\Delta \wb \over \wb}
\nonumber\\&& \quad
    -0.17 {\Delta \Omega_{\Lambda} \over \Omega_{\Lambda}}
    -1.1  {\Delta \Omegat \over \Omegat} \nonumber\\
    &\approx&
        -0.11 \Delta w - 0.48 {\Delta h \over h}
    +0.07 {\Delta \wb \over \wb}
    -0.15 {\Delta \Omega_{m} \over \Omega_{m}}
\nonumber\\&& \quad
    -1.4  {\Delta \Omegat \over \Omegat} \,,
    \label{eqn:lavar}
\end{eqnarray}
where the leading order dependence is on $\Omegat$ and $h$
\begin{eqnarray}
    {\Delta \ell_{\rm eq} \over \ell_{\rm eq}}
    &\approx&
    -0.11 \Delta w 
    +0.5 {\Delta \wm \over 
                           \wm}
    -0.17 {\Delta \Omega_{\Lambda} \over \Omega_{\Lambda}}
    -1.1  {\Delta \Omegat \over \Omegat} \nonumber\\
    &\approx&
      -0.11 \Delta w 
    + {\Delta h \over 
                           h}
    + 0.59 {\Delta \Omega_{m} \over \Omega_{m}}
    -1.4  {\Delta \Omegat 
    \over \Omegat} \,,
\end{eqnarray}
which depends more strongly on $\Omega_{m}$, and
\begin{eqnarray}
    {\Delta \ell_{D} \over \ell_{D}}
    &\approx&
    -0.11 \Delta w 
    -0.21 {\Delta \wm \over 
                           \wm}
    +0.20 {\Delta \wb \over \wb}
\nonumber\\&& \quad
    -0.17 {\Delta \Omega_{\Lambda} \over \Omega_{\Lambda}}
    -1.1  {\Delta \Omegat
    \over \Omegat} \nonumber\\
    &\approx&
        -0.11 \Delta w - 0.42 {\Delta h \over h}
    +0.20 {\Delta \wb \over \wb}
\nonumber\\&& \quad
    -0.12 {\Delta \Omega_{m} \over \Omega_{m}}
    -1.4  {\Delta \Omegat \over \Omegat} \,.
\end{eqnarray}
which depends more strongly on the baryon abundance $\wb$.
Note that the sensitivity to $\Omegat$ increases from the often quoted
$-0.5  \Delta\Omegat/\Omegat$ as $\Omega_\Lambda$ increases (\cite{Wei00} 2000;
M. Turner, private communication).

Ideally one would like to extract these three numbers and
the baryon-photon ratio $R_*$ directly from the data.  
The acoustic scale is readily extracted via the
position of the first and/or other higher peaks.  
The other quantities however
are less directly related to the observables. 
We instead choose to translate the parameter dependence
into the space of the observations: in particular
the height of the first three peaks.  

The height of the first peak 
\begin{equation}
    H_{1} \equiv \left({ \Delta T_{\ell_{1}} \over \Delta T_{10}}
		\right)^2\,.
\end{equation}
may be 
raised by increasing the radiation driving force (lowering 
$\ell_{\rm eq}$ or $\wm$) or the baryon drag (raising $\wb$).
However it may also be lowered by filling in the anisotropies
at $\ell \approx 10$ through the ISW 
effect (raising $\Omega_\Lambda$ or
$w$, or lowering $\Omegat$), reionization 
(raising the optical depth $\tau$), or
inclusion of tensors.  Each of the latter effects leaves the morphology
of the peaks essentially unchanged.  Because $H_1$ depends on
many effects, there is no simple fitting formula that describes
it. Around the $\Lambda$CDM model with $H_1=7.4$, it is crudely    
\begin{eqnarray}
    {\Delta H_{1} \over H_{1}} &\sim&
    -0.5 {\Delta \wm \over \wm} 
    +0.4 {\Delta \wb \over 0.02}
    -0.5 {\Delta\Omega_{\Lambda}}
    +0.7 {\Delta \Omegat}
\nonumber\\&& \quad
    2.5 {\Delta n} 
    -1 {\Delta\tau}
    -0.3 {\Delta w}
    - 0.76 {\Delta r \over 1+0.76r}\,.
\nonumber
\end{eqnarray}
where the tensor contribution
$r \equiv 1.4 (\Delta T_{10}^{\rm (T)}/\Delta T_{10}^{\rm (S)})^2$.
This scaling should only be used for qualitative purposes.

The height of the second peak relative to the first is written
\begin{equation}
H_{2} \equiv \left( \Delta T_{\ell_{2}} \over \Delta T_{\ell_{1}}
		\right)^2
	\approx 
	{ 0.925 \wm^{0.18} (2.4)^{n-1}\over 
	[1 + (\wb/0.0164)^{12 \wm^{0.52}}]^{1/5}}\,,
	\label{eqn:h2form}
\end{equation}
where $n$ is the scalar tilt and $\ell_2/\ell_1 \approx 2.4$.  
This approximation breaks down at high $\wb$ and $\wm$ as the
second peak disappears altogether.
In the $\Lambda$CDM model with $n=1$, $H_2 = 0.51$ and parameter
variations yield 
\begin{equation}
{\Delta H_{2} \over H_{2}} \approx 0.88 \Delta n
-0.64 {\Delta \wb \over \wb}
+0.14 {\Delta \wm \over \wm}\,.
\end{equation}
The effect of tilt is obvious.  Baryons lower $H_2$ by increasing
the modulation that raises all odd peaks.  The dependence on the
matter comes from two competing effects which nearly cancel around
the $\Lambda$CDM: increasing $\wm$ (lowering $\ell_{\rm eq}$) 
decreases the radiation driving
and increases $H_2$ but also increases the depth of potential
wells and hence the modulation that lowers $H_2$.

For the third peak, these effects add rather than cancel.  When
scaled to the height of the first peak, which is also increased
by raising the baryon density, the $\Omega_b h^2$ 
dependence weakens leaving 
a strong dependence on the matter density
\begin{eqnarray}
H_{3} &\equiv& \left( \Delta T_{\ell_{3}} \over \Delta T_{\ell_{1}}
		\right)^2 
\nonumber\\
       & \approx &
        2.17\left[ 1 + \left( {\wb \over 0.044} \right)^{2}\right]^{-1}
        \wm^{0.59}
        (3.6)^{n-1}
\nonumber\\
 	&& \quad \times\left[1 + 1.63\left(1-{\wb \over 0.071}\right)\wm
	\right]^{-1}
\end{eqnarray}
where $\ell_3/\ell_1 \approx 3.6$. Around the fiducial $\Lambda$CDM
model where $H_3=0.50$ 
\begin{equation}
{\Delta H_{3} \over H_{3}} \approx 0.41 \Delta n
-0.31 {\Delta \wb \over \wb}
+0.53 {\Delta \wm \over \wm}\,.
\end{equation}

We emphasize that the phenomenology in terms of $\ell_A$,
$\ell_{\rm eq}$ and $\ell_D$ is relatively robust, predictive
of morphology beyond the first three peaks and readily
generalizable to models outside the adiabatic
cold dark matter paradigm.  The specific scalings of $H_1$,
$H_2$ and $H_3$ with cosmological parameters are only valid
within the family of adiabatic CDM models.  Furthermore, as the data continue
to improve, the fits must also be improved from their current
few percent level accuracy.  The number of phenomenological
parameters must also increase to include at least both the 
heights of the peaks and the depths of the troughs for 
all observed peaks.

\clearpage

\end{document}